# THE DEFINABILITY OF FIELDS

D.J. BENDANIEL
The Johnson School, Cornell University
Ithaca, NY  14853

**Abstract.**  We look for a deep connection between mathematics and physics.  Our approach is to propose a set theory T which leads to a concise mathematical description of physical fields and to a finite unit of action.  The concept of "definability" of fields is then introduced.  Definability of fields in T is necessary and sufficient for quantization and sufficient to avoid physical antinomies.

Consider a most interesting philosophical question:  why is mathematics so useful in describing physics?  Physics is based solely on observations.  Mathematics, on the other hand, is based solely on deductive reasoning from abstract axioms, with no appeal to observations whatsoever.  Why, then, should mathematics be in any way related to physics?

In 1960, Wigner [1] published a charming paper, in which he gave several examples of mathematics being "unreasonably effective" in describing nature, the inference being there must actually be some deep connection.  Our approach is to propose an axiomatic foundation for physics closely related to that which has been developed for mathematics.  The connection can then explained and, moreover, the physical universe can be viewed as entirely governed by a formal system.

The usual foundation of mathematics is the set theory of Zermelo-Fraenkel (ZF).  We shall start by removing from ZF the axiom schema of replacement (AR).  The reason why this axiom is left out is that without it, or specifically without a certain part of it, non-denumerable infinities cannot be derived.  Now, the axiom schema of replacement is composed of two axioms which can be considered separately as the axiom schema of subsets [2] and the axiom schema of bijective replacement (ABR). The axiom schema of subsets is the part required for non-denumerable infinities. To remove these infinities, we must delete the axiom schema of subsets from ZF by requiring replacement to be bijective, forming ZF-AR+ABR (See appendix).

In order to understand this foundation, we must give a close look at the axiom of infinity. The axiom of infinity is naively viewed as asserting the existence of the infinite set, usually called $\omega$, containing only all the finite natural numbers. More generally, this axiom asserts the existence of infinite sets $\omega^*$ which also contain infinite natural numbers. To obtain $\omega$ in ZF, we must first use the axiom schema of subsets to establish the existence of the set of all the sets created by the axiom of infinity and then, by the axiom of regularity, show this set of sets has a minimal element. That minimal element is $\omega$. Holmes [3] showed that no model of ZF–AR+ABR contains $\omega$. This confirms that the axiom schema of subsets is not hidden in the other axioms and implies that ZF–AR+ABR is uniformly dependent on $\omega^*$, that is, every theorem holds for any $\omega^*$. We shall now refer to all the members of any $\omega^*$ as "integers". Infinite integers are defined as those members of any $\omega^*$ mapping one-to-one with any $\omega^*$ and therefore *are* any $\omega^*$. Finite integers are those not mapping one-to-one with any $\omega^*$ and are designated as *i, j, k, ℓ, m, M, n,* or N.

To ZF–AR+ABR we can adjoin "all sets are constructible". By constructible sets we mean sets which are generated sequentially by a process, one after the other, such that the process well-orders the sets. We shall call this theory T. Goedel [4] showed an axiom of constructibility could be added consistently to ZF giving a system usually designated as ZF+. We are, therefore, on safe ground since T is a sub-theory of ZF+.

Some important theorems of T are quite different from ZF. There can be no inductive proofs from the finite to the infinite such as are possible in ZF. In addition, in T all sets of finite integers are finite, unlike in ZF where we can have infinite sets of finite integers. On the other hand, any set of integers in T which is infinite (and we do have such sets) must contain both finite *and infinite* integers.

We can now derive real numbers using "non-standard" methods [5]. First, the usual definition of "rational numbers" is as the ratio of two finite integers. We can here define the set of ratios of two integers as an "enlargement" of the rational numbers. The "non-standard reals" are then contained within this enlargement or we can say that the non-standard reals *are* this enlargement, since the set of all rationals does not exist in T, just as the set of all finite integers does not exist. An "infinitesimal" is a non-standard real which is "equivalent" to 0, that is, letting $x$ signify a non-standard real and using the symbol "=" to signify equivalence, $x = 0 \leftrightarrow \forall k[x < 1/k]$. Any non-standard real is either equivalent to 0 or defined as "finite", that is, $x \neq 0 \leftrightarrow \exists k[1/k \leq x]$. Now we have to be careful of what we mean by equality. Non-standard reals which do not differ we can call "identical" and we shall use the symbol "≡". Identical non-standard reals are, of course, also equivalent. Thus, two non-standard reals in this theory are either "equal" (that is, equivalent) or their difference must be finite. With this notion of equality, we can create the "real numbers" as a set whose elements are the integers times a fixed infinitesimal. The reals are therefore a subset of the non-standard reals.

An "equivalence-preserving" bijective mapping $f(x,u)$ between the non-standard reals of finite intervals X and U, where $x \in X$ and $u \in U$, is a "function of real variables": $\forall x_1,x_2,u_1,u_2 \; f(x_1,u_1) \wedge f(x_2,u_2) \to (x_1 = x_2 \leftrightarrow u_1 = u_2)]$. These functions are biunique and continuous. A calculus restricted to these functions can be developed in T just as in ZF, since the axiom of subsets is a theorem of T for the special case of bijective mappings. Non-biunique functions of real variables must be built up by attaching biunique pieces. In general, $f_1(x,u) \vee f_2(x,u)$ is a function of real variables with domain $X_1 \cup X_2$ and range $U_1 \cup U_2$ if $\forall x_1, x_2, u_1, u_2 [f_1(x_1, u_1) \vee f_2(x_1, u_1) \wedge f_1(x_2, u_2) \vee f_2(x_2, u_2) \rightarrow (x_1 \equiv x_2 \rightarrow u_1 \equiv u_2)]$ where $f_1(x,u)$ and $f_2(x,u)$ are both functions of real variables. This necessitates that *all functions of real variables are of bounded variation*.

We've seen these functions before. They are familiar to mathematical physicists as those functions of real variables which are uniformly convergent with sums of eigenfunctions of the Sturm-Liouville problem:

For an irreducible biunique eigenfunction piece whose end points are a, b, where $a \neq b$, and $u \dfrac{du}{dx} = 0$ at these endpoints,

$$\int_a^b [p\left(\frac{du}{dx}\right)^2 - qu^2]dx = \lambda \int_a^b ru^2 dx \qquad [1]$$

and where p, q and r are functions of real variables.

In order to obtain the basic biunique pieces from which to build up these eigenfunctions, we use the variational method: $\lambda$ is minimum for $\int_a^b ru^2 dx$ constant. Therefore, all functions of real variables in T can in principle be built up just from sums of these basic biunique eigenfunction pieces, every piece determined from the same integral expression.

These pieces, in turn, provide a concise description of physical phenomena. Consider biunique eigenfunction pieces $u_1$ and $u_2$, where $u_1$ is a function of $x_1$ and $u_2$ is a function of $x_2$. Then the condition $\lambda_1 - \lambda_2 = 0$ leads directly to an integral over $dx_1 dx_2$ which vanishes. If we let $x_1$ be "space" and $x_2$ be "time", then the integrand is the Lagrange density of a "wave field" $u_1 u_2$. Moreover, with very little work, we can extend this result to generalized vector wave fields in finitely many space-like (i) and time-like dimensions (j) (e.g., strings).

$$0 = \sum_i \lambda_i - \sum_j \lambda_j$$

$$0 = \sum_i \lambda_i \int \underline{\Psi} \bullet \underline{\Psi} \, d\tau - \sum_j \lambda_j \int \underline{\Psi} \bullet \underline{\Psi} \, d\tau \qquad [2]$$

where

$$\underline{\Psi} = \sum_\ell \Psi_\ell \underline{i}_\ell, \quad \Psi_\ell = \prod_i u_{\ell i} \prod_j u_{\ell j}, \text{ and } d\tau = \prod_i r_i dx_i \prod_j r_j dx_j,$$

giving

$$0 = \int \left\{ \sum_i \frac{1}{r_i} \left[ p_i \left( \frac{\partial \underline{\Psi}}{\partial x_i} \bullet \frac{\partial \underline{\Psi}}{\partial x_i} \right) - q_i \underline{\Psi} \bullet \underline{\Psi} \right] - \sum_j \frac{1}{r_j} \left[ p_j \left( \frac{\partial \underline{\Psi}}{\partial x_j} \bullet \frac{\partial \underline{\Psi}}{\partial x_j} \right) - q_j \underline{\Psi} \bullet \underline{\Psi} \right] \right\} d\tau$$

This integral, over each irreducible element of the field composed of pieces in finitely many space-time dimensions, has equal and opposite potential and kinetic parts.

Let the product of $\int \underline{\Psi} \bullet \underline{\Psi} d\tau$ with $\sum_i \lambda_i$ or $\sum_j \lambda_j$ be called "action" and symbolized by $\hat{\alpha}$. Here is a proof in T that there must be a finite indivisible unit of action:

I. The field $\underline{\Psi}$ is either non-existent (in which case $\hat{\alpha} \equiv 0$) or is a function of real variables (in which case $\hat{\alpha} \neq 0$). Thus, $\hat{\alpha} = 0 \leftrightarrow \hat{\alpha} \equiv 0$.

II. Since action is a real number, then $\hat{\alpha} = z\hat{h}$, where the constant $\hat{h} \not\equiv 0$ and z is any integer (thus, $z = 0 \leftrightarrow z \equiv 0$).

III. Then, $\hat{\alpha} \equiv 0 \rightarrow z \equiv 0$, since $\hat{h} \not\equiv 0$.

IV. Accordingly, $z \neq 0 \rightarrow z \not\equiv 0 \rightarrow \hat{\alpha} \not\equiv 0 \rightarrow \hat{\alpha} \neq 0 \rightarrow \hat{h} \neq 0$, $\therefore \hat{h}$ is a finite unit of action.

For simplicity, we consider only a string periodic in one spatial dimension and one time dimension and we shall set both the spatial and time periods at unity. We can define "energy" as the number of units of action per time period. Then, the energy associated with the m$^{th}$ eigenstate occurs only in quanta of $4\hat{h}m$. Total energy in all the eigenstates is conserved, furthermore, if and only if the total number of units of action is the same in any distribution of energy among eigenstates.

We have set the stage to introduce a useful concept: the definability of a physical field in the theory T. Every set (in T) of finite integers is finite and therefore ipso facto definable. We can now say that a field is "definable in T" if the set of all possible distributions of energy among eigenstates (which, of course, is the set which underlies quantum statistics) can be arithmetized and mirrored by some set (in T) of finite integers. To understand this, we need only look again at the simple string. Every set of amplitudes $j_m$ such that total energy is conserved can be represented uniquely by a finite integer:

$$\left\{ j_m \mid \sum_m^M j_m m = M \right\} \Rightarrow \prod_m^M (P_m)^{j_m} \quad [3]$$

where $P_m$ is the m$^{th}$ prime starting with 2. We can then form the set of all such finite integers. Thus, if the total energy is finite, quantization is sufficient for definability. To investigate necessity, furthermore, we need only look to the correspondence principle: if the energy is finite and $\hat{h}$ were to go to 0, then the integer M must go to infinity. In that case, the set of all possible distributions of energy among eigenstates can no longer be mirrored by any set (in T) of finite integers. Therefore, definability of a field in T is both necessary and sufficient for quantization. This argument can also be generalized to vector wave fields in finitely many space-time dimensions.

Finally, we can discuss the philosophical meaning of "definability in T" and suggest a connection between the foundations of mathematics and physics. In ZF there is a non-denumerable infinity of sets of finite integers. In general, a set U of finite integers is definable in any set theory if there exists a formula $\Phi(n)$ from which we can unequivocally determine whether or not a given finite integer n is in the set or not. Since there are only denumerably many defining formulae, there can be only denumerably many definable sets of finite integers. It follows that almost all sets of finite integers in ZF are not definable. When a set of finite integers is not definable, then there will be at least one finite integer for which it is not possible to determine whether it is in the set or not. In that case, we will obtain within the theory an antinomy of the form $\exists n\, n \in U \leftrightarrow n \notin U$. Now, we can make our deep connection with physics by asserting that *physical fields cannot have antinomies*. If otherwise, the universe would not operate. It would stop dead or tear apart. Or, more precisely, a field whose set of all energy distributions would be mirrored by an undefinable set of finite integers could not exhibit causality. One could ask, "Can we not just restrict ourselves to the definable sets of finite integers in ZF?" The answer is that, when we create the set of all definable sets of finite integers in ZF, we find that set itself to be undefinable; we can never decide on at least one of its members, a result due to Tarski [6].

Thus, to avoid physical antinomies, we go to a foundation in which all sets of finite integers are finite. The foundation T, while less rich than ZF, is just rich enough to contain quantized physical fields and, as we have shown, these fields are definable.

In conclusion, we have made a deep connection between a certain set-theoretical foundation which we have here called T and physical fields. That is, these fields appear to obey all the rules of the set theory, in their functional form, quantization and causality.

# Appendix

Extensionality- Two sets with just the same members are equal.

$$\forall x \forall y [\forall z (z \in x \leftrightarrow z \in y) \rightarrow x = y]$$

Pairs- For any two sets, there is a set which contains just them.

$$\forall x \forall y \exists z [\forall w\, w \in z \leftrightarrow w = x \vee w = y]$$

Union- For any set of sets, there is a set with just all their members.

$$\forall x \exists y \forall z [z \in y \leftrightarrow \exists u (z \in u \wedge u \in x)]$$

Infinity- There is a set with members determined in endless succession.

$$\exists x [\exists y\, y \in x \wedge \forall y [y \in x \rightarrow \exists z [z \in x \wedge z \neq y \wedge \forall u (u \in y \rightarrow u \in z)]]]$$

Power Set- For any set, there is a set containing just all its subsets.

$$\forall x \exists P(x) \forall z [z \in P(x) \leftrightarrow \forall u (u \in z \rightarrow u \in x)]$$

Regularity- Every set has a minimal member.

$$\forall x [\exists y\, y \in x \rightarrow y [\exists y \in x \wedge \forall z \neg (z \in x \wedge z \in y)]]$$

Schema of Bijective Replacement- For any set, replacing its members one-for-one with members from some set creates a set.

Let $\phi(s,t)$ be any formula in which (s,t) is free,

$$\forall x \in z \exists y \in w [\phi(x,y) \wedge \forall u \in z \forall v \in w [\phi(u,v) \rightarrow u = x \leftrightarrow y = v]] \rightarrow$$

$$\exists r \forall t \in w [t \in r \leftrightarrow \exists s \in z\, \phi(s,t)]$$